\documentclass[apsrev,superscriptaddress,showpacs,showkeys]{revtex4}

\begin{document}
\title{Conserved Correlation in $\cal{PT}$-symmetric Systems: Scattering and Bound States}
\author{Kumar Abhinav}
\email{kumarabhinav@iiserkol.ac.in}
\affiliation{Indian Institute of Science Education and Research-Kolkata, Mohanpur Campus, Nadia-741252, West Bengal, India}
\author{Arun Jayannavar}
\email{jayan@iopb.res.in}
\affiliation{Institute of Physics, Bhubaneswar-751005, Orissa, India}
\author{P. K. Panigrahi}
\email{pprasanta@iiserkol.ac.in}
\affiliation{Indian Institute of Science Education and Research-Kolkata, Mohanpur Campus, Nadia-741252, West Bengal, India}

\begin{abstract} 
For one-dimensional $\cal{PT}$-symmetric systems, it is observed that the {\it non-local} product
$\psi^*(-x,t)\psi(x,t)$, obtained from the continuity equation can be interpreted as a conserved correlation function. 
This leads to physical conclusions, regarding both discrete and continuum states of such
systems. Asymptotic states are shown to have necessarily broken $\cal{PT}$-symmetry, leading to
modified scattering and transfer matrices. This yields restricted boundary conditions, {\it e.g.}, incidence
from both sides, analogous to that of the proposed $\cal{PT}$ CPA laser \cite{Longhi}. The interpretation
of `left' and `right' states leads to a \textit{Hermitian} $S$-matrix, resulting in the non-conservation of the
`flux'. This further satisfies a `duality' condition, identical to the optical analogues \cite{Dual}.
However, the non-local conserved scalar implements alternate boundary conditions in terms of `in' and `out'
states, leading to the pseudo-Hermiticity condition in terms of the scattering matrix. Interestingly, when
$\cal{PT}$-symmetry is preserved, it leads to stationary states with real energy, naturally interpretable
as bound states. The broken $\cal{PT}$-symmetric phase is also captured by this correlation, with
complex-conjugate pair of energies, interpreted as resonances.                 
\end{abstract}



\maketitle

\section*{Introduction}
A number of non-Hermitian Hamiltonians are known to have real spectra for certain range of parameter values.
In a different parameter regime, there exist complex conjugate pairs of energy, owing to their inherent
parity-time ($\cal{PT}$)-symmetry \cite{Bender}. Experimental realization of $\cal{PT}$-symmetric
optical systems \cite{Guo} has prompted several proposals \cite{Cao}, one of which is
the proposed Coherent Perfect Absorber (CPA), with only time reversal ($\cal{T}$) symmetry. It
shows perfect absorption of two laser beams, incident from two opposite directions, with definite phase and amplitude
relationships. This has been further generalized by involving non-linear systems \cite{Longhi}, wherein the
CPA emerges as a special case of the $\cal{PT}$ CPA laser. This system can exhibit 
spontaneous emission of two laser beams in opposite directions, physical realization of which
can be useful as very sensitive optical switches and sensors \cite{Longhi, Cao}. Very recently, suggestion that
CPA can be realized without $\cal{PT}$-symmetry has been made \cite{1most}, although the $\cal{PT}$-symmetric version
is still useful to understand the quantum mechanical analogues. 
\paragraph*{} Optical systems and their quantum mechanical counterparts differ at a fundamental level,
in the sense that, conditions like square integrability and the nature of the Hilbert space, comes into 
play in the later case. Existence of a well-defined inner-product is an additional necessity
for quantum mechanical systems. As an example, the  $\cal{PT}$ CPA laser can be studied in terms of
various matrix elements, corresponding to reflection and transmission coefficients, which is not possible 
for the quantum mechanical analogue. The absence of a positive semi-definite inner-product for $\cal{PT}$-symmetric
systems, under the usual Dirac-von-Neumann construction of Hilbert space, has led to the redefinitions
of the same \cite{Norm}. Mostafazadeh showed that \cite{Most1}, under certain conditions, these Hamiltonians can be
pseudo-Hermitian, spanned on a bi-orthonormal basis. A complete prescription to obtain a
positive semi-definite inner-product for pseudo-Hermitian systems was finally given by Das and Greenwood \cite{Das}.
However, a general proof of equivalence of pseudo-Hermiticity and $\cal{PT}$-symmetry is still 
lacking, even for bounded spectrum-generating operators, as has been shown for more general 
cases \cite{Siegl}. 
\paragraph*{•}The anti-linear nature of the time-reversal operation is the root of the difficulty
in constructing a $L_2$ norm for $\cal{PT}$-symmetric systems. The corresponding anti-unitary evolution
of the system prevents the existence of a dual vector space, necessary for constructing a Hilbert space,
with a positive semi-definite norm leading to the quantum mechanical probability density. However, as parity
is an unambiguous discrete symmetry in one-dimension, it is possible to generate a 
conserved `\textit{scalar product}', bypassing the inconvenience due to anti-linearity. 
This is not the case for one-dimensional systems, obtained from higher-dimensional systems under symmetry
reduction. Further, the aforementioned scalar product does not correspond
to a conserved probability, which interestingly can be interpreted as a conserved correlation, given by
$\psi^*(-x)\psi(x)$ \cite{BQZ}. It is obtained through the
use of equation of motion for $\cal{PT}$-symmetric systems, connecting two parity-opposite spatial locations.
This non-local correlation, when integrated over all space, yields a conserved charge of the theory. This
explains somewhat different asymptotic behavior reported in Ref.\cite{Guo} and proposed in Refs.\cite{Longhi,Cao}.
\paragraph*{}In the present paper, we systematically explore the implications of this non-local correlation in 
$\cal{PT}$-symmetric systems. In case of scattering \cite{Scattering}, this scalar can be viewed as
a correlation between states at two asymptotes ({$x=\pm\infty$), requiring non-local boundary conditions. 
The time-evolution of the system needs to be of the form $\exp(-iHt)$, with the Hamiltonian being complex.
It is observed that $\cal{PT}$-symmetry is necessarily broken for asymptotic states. The real energy phase (unbroken $\cal{PT}$-symmetry)
of these systems corresponds to stationarity of the correlation scalar, with the aforementioned
temporal exponent being unitary. Here, the eigenfunctions are stationary, with discrete eigenvalues, as
evaluated directly in numerous examples \cite{Discrete}, admitting `bound state' interpretation. The broken
$\cal{PT}$-symmetric phase has also been captured with complex-conjugate pairs of energies \cite{Bender},
with corresponding eigenfunctions related through $\cal{PT}$-transformation. The spatial part of the `current' is not
conserved in this case, due to the presence of gain/loss, which can be interpreted as resonance.
\paragraph*{}The paper has been organized as follows. In Sec.I, we study a generic 1-D quantum mechanical
$\cal{PT}$-symmetric system, wherein the corresponding equation of motion is utilized to arrive at the conserved
non-local scalar. Its implication towards the norm for the $\cal{PT}$-symmetric
systems is pointed out. Further, the symmetry structure of the scattering process is shown to be
different from that of Hermitian systems. New conditions are shown to be satisfied by the $S$-matrix,
with pseudo-Hermiticity being achieved through the incorporation of non-locality into the boundary conditions. 
In Sec.II, the properties of the wave-functions of a $\cal{PT}$-symmetric system through the continuity
equation are analyzed. Stationary states are shown to have real eigenvalues and unitary temporal evolution,
with $\cal{PT}$-symmetry being necessarily preserved. It is also shown that spontaneous breaking of
$\cal{PT}$-symmetry leads to complex-conjugate pairs of eigenvalues, with corresponding eigenfunctions
related through $\cal{PT}$-transformation. The $\cal{PT}$-symmetric boundary conditions for
scattering are obtained, which are more constrained. Transmission, and also complete absorption, are possible,
only if plane-waves are incident from both directions, analogous to the observations reported in Ref. \cite{Longhi}. 
In Sec.III, for the purpose of demonstration, we analyze the \textit{complexified} 1-D Scarf-II potential, which
is $\cal{PT}$-symmetric. It is asymptotically constant, yielding scattering states which are plane-waves.
The corresponding probability flux is not conserved, under the Hermitian norm. The asymptotic
coefficients are shown to satisfy the non-local boundary conditions, following the non-local conserved scalar.
Finally, we conclude with remarks on possible implications of our results and subsequent uses.

\section{Continuity equation for $\cal{PT}$-symmetric systems: Implication for the $S$-matrix}

\subsection*{The non-local scalar and the norm}
As is known, operator action in quantum mechanics can be defined without the help of a well-defined norm
\cite{LL}, so long as expectation values are not summoned into the picture, given a right-operation (or left,
but obviously not both) is defined. Although the matrix elements can be evaluated only after fixing a norm,
algebraic conditions can still be obtained, from the equation of motion. For a $\cal{PT}$-symmetric system,
the 1-D Schr\"odinger equation: 
$-\frac{\hbar^2}{2m}\frac{{\partial}^2}{\partial x^2}\psi(x,t)+V(x)\psi(x,t)=i\hbar \frac{\partial}{\partial t}\psi(x,t)$,
does not lead to the usual definition of the probability current. In the Hermitian case, we arrive at the
equation of continuity by using the Schr\"odinger equation, together with its complex-conjugate counterpart.
If the potential is $\cal{PT}$-symmetric, one needs to take a
$\cal{PT}$-transformation of the equation, in conjunction with complex-conjugation \cite{BQZ}, in order to
obtain the equation of continuity:

\begin{equation}
\frac{\hbar}{2im}\frac{\partial}{\partial x}\left(\psi(x,t) \frac{\partial}{\partial x} {\psi}^{*}(-x,t)-{\psi}^{*}(-x,t)\frac{\partial}{\partial x} \psi(x,t)\right)=\frac{\partial}{\partial t}\left({\psi}^{*}(-x,t)\psi(x,t)\right).\label{cont}
\end{equation}
Thus, one arrives at a new definition of flux, which is conserved. This is achieved at the expense of a real positive-definite norm of the 
Hermitian theory, as is evident from the time-derivative part of the above equation. This leads to re-interpretation of the scattering 
process. It is evident from the above equation that, the scalar that naturally emerges from the system dynamics is neither local
nor real; and hence, cannot be interpreted as probability density in the line of Hermitian systems. Instead, it is more suitable to be 
identified as a correlation function between two parity-opposite spatial points. This is physically meaningful, as a complex potential
can lead to `change of state' through emission or absorption. However, upon integration,
it does yield a conserved scalar of the $\cal{PT}$-symmetric system, which suggests towards a modified norm \cite {Bender}.
Furthermore, on identifying $\psi(x,t){\psi}^{*}(-x,t)=\psi(x,t)PT\psi(x,t)$, the general, non-local,
$\cal{PT}$-symmetric scalar product between two distinct wave-functions $\phi(x,t)$ and $\psi(x,t)$ can be defined as,

\begin{eqnarray}
\int_{-\infty}^{\infty}\phi(x,t)PT\psi(x,t)dx&=&\int_{-\infty}^{\infty}\phi(x,t)\psi^{*}(-x,t)dx\nonumber \\
&=&\int_{-\infty}^{\infty}\psi^{*}(x,t)\phi(-x,t)dx\nonumber \\
&=&\int_{-\infty}^{\infty}\psi^{*}(x,t)P\phi(x,t)dx.\label{norm2}
\end{eqnarray}
The last result appears as a generalization of the Dirac-von Neuman scalar product, which has already been proposed \cite{NBM}.
The exchange of $\phi$ and $\psi$ is due to parity, which is well-defined in one dimension (1-D). This interchange is in the spirit 
of anti-unitary operation $|\alpha^{\star}\rangle=\Theta|\alpha\rangle$ \cite{Sakurai}, leading to $\langle\alpha^{\star}|\beta^{\star}\rangle=\langle\beta|\alpha\rangle$. The anti-unitary operator $\Theta$ is a generalization of the anti-linear operator $T$. 
\paragraph*{}On the other hand, a $\cal{PT}$-symmetric Hamiltonian, when treated similar to a pseudo-Hermitian
one \cite{Das}, leads to,

\begin{eqnarray}
H&=&(PT)H(PT)^{-1}\equiv PTHPT=P(THT)P=PH^{*}P\nonumber\\
\text{or,}\nonumber\\
H&=&P(H^{\dagger})^{\tau}P=P\tau H^{\dagger}{\tau}^{-1}P=(P\tau)H^{\dagger}(P\tau)^{-1},\label{pseudo}
\end{eqnarray} 
where we define $\tau$ as the \textit{transposition operator}, relating a particular matrix to its transpose through similarity 
transformation. This depends on the particular matrix and its representation in the basis of choice and preserves the anti-linear
nature of the time-reversal operator. This is identical to the definition of pseudo-Hermitian Hamiltonian, $H=\eta^{-1}H^{\dagger}\eta$ 
\cite{Das}, for $(P\tau)^{-1}=\eta$. Here, $\tau$ implies transposition only for the Hamiltonian operator, hence,

\begin{equation}
\langle \phi|(P\tau)^{-1}|\psi\rangle=\langle\phi|\tau^{-1}P|\psi\rangle=\langle\phi|\tau P|\psi\rangle,\label{norm1}
\end{equation}
as transposition is idempotent. Using the Schr\"odinger representation, the relation $THT=H^*$ can be realized as:

\begin{equation}
\langle m|THT|n \rangle=\int_{-\infty}^{\infty}dx\langle m|x\rangle H(-x)\langle x|n\rangle.
\end{equation}
As for a $\cal{PT}$-symmetric Hamiltonian, $H(-x)\equiv H^*(x)$, the last term of the above equation is $<m|H^*|n>$.
\paragraph*{}It is evident that, 
construction of a pseudo-Hermitian norm for $\cal{PT}$-symmetric systems, necessarily incorporates anti-linearity through $\tau$. 
That such an operator is representation-dependent is physically justified, as the form of $\eta$ always depends on the pseudo-Hermitian 
system itself. However, the fact that transposition necessarily requires a predefined scalar product, actually makes the norm in the 
second prescription ill-defined. The first prescription yields a well-defined conserved scalar product, however it does not qualify 
as the norm, as positive definiteness is not ensured. Also, which state is to be chosen for right-operation is not clear if one naively 
starts with this prescription, which further emphasizes the inherent non-locality.
\paragraph*{}These inadequacies extend to the earlier difficulty for calculating the scattered `flux' for a 
$\cal{PT}$-symmetric system. There have been prescriptions to make the above conserved scalar product 
positive-definite \cite{NBM}, for systems with finite Hilbert spaces. Unbounded systems are yet 
to be tackled, not to mention the already stated difficulty of generic bounded spectral operators \cite{Siegl}. In case of asymptotically 
Hermitian systems, the second prescription appears more suitable of the two, as it requires 
generalization of $\tau P$ to obtain a proper pseudo-Hermitian norm, corresponding to $\eta(x\rightarrow\pm\infty)\longrightarrow I$. 

\subsection*{The scattering properties}
The above conserved correlation imposes novel boundary conditions for $\cal{PT}$-symmetric
systems. They impose additional constraints on the system than their Hermitian counterparts, yielding unique algebraic structure
and clear distinctions between bound, resonance and asymptotic states. Analysis of the generic scattering by such systems enables
one to obtain the same. For comparative clarity, we consider a generic
one-dimensional $\cal{PT}$-symmetric potential, which is asymptotically Hermitian (converges to a unique real
constant as $x\longrightarrow \pm \infty$), admitting scattering states which are plane-waves. Then the general asymptotic
solution can be written as,

\begin{equation}
\psi(x)\longrightarrow \left\{\begin{array}{rl}{Ae^{ikx}+Be^{-ikx},}~~~ & \text{when}~x\longrightarrow -\infty, \\
{Ce^{ikx}+De^{-ikx},}~~~ & \text{when}~x\longrightarrow \infty, \end{array} \right.\label{Asym}
\end{equation}
with $A,B,C,D$ being complex (C) numbers.
\paragraph*{•} For a Hermitian potential which is asymptotically well-behaved, the asymptotic co-efficients are linked as,

\begin{eqnarray}
\left(
\begin{array}{c} C\\D
\end{array} \right)=M\left(
\begin{array}{c} A\\B
\end{array} \right)~~~\text{and}~~~
\left(
\begin{array}{c} B\\C
\end{array} \right)=S\left(
\begin{array}{c} A\\D
\end{array} \right),\label{def}
\end{eqnarray}
where $M$ and $S$ are transfer and scattering matrices respectively, linking left-right and incoming-outgoing states.
In the Hermitian case, the form of the conserved current: $j(x,t)=\frac{\hbar}{2im}\left[{\psi}^*(x,t)\frac{\partial}
{\partial x}\psi (x,t)-\psi (x,t)\frac{\partial}{\partial x}{\psi}^*(x,t)\right]$, leads to the unitarity of the S-matrix,
and transfer matrix satisfies the condition,

\begin{equation}
M^{\dagger}\left(
\begin{array}{cc} 1 & 0\\0 & -1
\end{array} \right)M=\left(
\begin{array}{cc} 1 & 0\\0 & -1
\end{array} \right).\label{su2}
\end{equation}
It is crucial to note that, though the notions of `incoming' and `outgoing' are represented by the pairs ($A,D$) and ($B,C$) 
respectively, the quantum state $\psi$ appearing in the expression for probability flux can also be classified as `left-asymptotic' (at $-\infty$) 
and `right-asymptotic' (at $\infty$), represented by pairs ($A,B$) and ($C,D$) respectively \cite{Merzbacher}. We follow both the
notions for the $\cal{PT}$-symmetric case.
\paragraph*{\bf The left-right interpretation:}Following the left-right asymptotic state convention, {\it i.e.}, $\psi_L=Ae^{ikx}+Be^{-ikx}$
and $\psi_R=Ce^{ikx}+De^{-ikx}$  respectively, conservation of the $\cal{PT}$-symmetric `current' leads to, 

\begin{equation}
AB^*-BA^*=CD^*-DC^*.\label{flux}
\end{equation}
The transfer matrix then obeys,

\begin{equation}
M^{\dagger}\left(
\begin{array}{cc} 0 & -1\\1 & 0
\end{array} \right)M=\left(
\begin{array}{cc} 0 & -1\\1 & 0
\end{array} \right).\label{su11}
\end{equation}
Further, the $S$-matrix is {\it Hermitian}, instead of being unitary. This is not surprising, since
a conserved `probability flux' cannot be constructed under the standard prescription of scalar 
product, which requires the $S$-matrix to be unitary. Further, a localized flux cannot be interpreted from Eq.\ref{flux}, which
can be attributed to the non-local character of the `charge' $\psi^*(-x)\psi(x)$. Despite of this fact, unique additional conditions 
on scattering states will be concluded in the following section, which are necessary for explaining known physical cases. We would like
to add  that, the notion of Hermitian conjugation used here is purely mathematical. Hermitian conjugate of any matrix $\Lambda$ 
is taken to be the matrix that results into the dual of any vector $Y=\Lambda X$ by left-operating on the dual of vector $X$. The
structure of Eq.\ref{flux} allows this construction, and there is no attempt to extract any physical interpretation for this 
Hermitian conjugation, unlike in usual quantum mechanics. But even then, the defining meanings of $M$ and $S$ holds, owing 
to boundary conditions. A re-defined physical norm must only affect the elements of these matrices, but not their definitions.
\paragraph*{•} As mentioned earlier, a $\cal{PT}$-symmetric potential has the general form,

\begin{equation}
V(x)=V_{even}(x)+iV_{odd}(x),\label{pt}
\end{equation}
where the suffixes mention respective parity of the functional parts of the potential, which are \textit{real}. Then, clearly, 
$H^*\left(V_{odd}(x)\right)=H\left(-V_{odd}(x)\right)$. Let ${\psi}_{\pm}(x,t)$ be solutions to $H\left(\pm V_{odd}(x)\right)$. 
The corresponding S-matrices, $S_{\pm}$, are Hermitian. On considering the time-independent scenario, if $V_{odd}(x\rightarrow\infty)\longrightarrow 0$ 
and if $V_{even}(x)$ is a constant asymptotically, the asymptotic momenta $k_{\pm}$ will be related as $k^{*}_{\pm}=k_{\mp}$. 
It is clearly seen that, ${\psi}^{*}_{+}(x,t)$ and ${\psi}_{-}(x,t)$ are the eigenfunctions to $H\left(-V_{odd}(x)\right)$, 
whereas ${\psi}^{*}_{-}(x,t)$ and ${\psi}_{+}(x,t)$ are the eigenfunctions to $H\left(V_{odd}(x)\right)$ at the two asymptotes.
As both the Hamiltonians asymptotically converge to that of a free particle, these solutions must be the same, 
as there is no degeneracy in the 1-D case. Same can be argued about the corresponding eigenvalues; the asymptotic coefficients 
for both the systems then satisfy,

\begin{eqnarray*}
A_{+}^{*}=B_{-},~~~ B_{+}^{*}=A_{-}, \nonumber\\ 
C_{+}^{*}=D_{-}~~\text{and}~~G_{+}^{*}=F_{-}.
\end{eqnarray*}
The definition of S-matrix leads to,

\begin{eqnarray}
\left(\begin{array}{cc} B_{+}^{*} & C_{+}^{*}\\ \end{array}\right)&=&\left(\begin{array}{cc} A_{+}^{*} & D_{+}^{*}\\ \end{array}\right)S^{\dagger}(V_{odd})\nonumber\\
\text{or,}\hspace{0.2in}\left(\begin{array}{cc} B_{+}^{*} & C_{+}^{*}\\ \end{array}\right)\left(\begin{array}{c} B_{-}\\C_{-} \end{array} \right)&=&\left(\begin{array}{cc} A_{+}^{*} & D_{+}^{*}\\ \end{array}\right)S^{\dagger}(V_{odd})\left(\begin{array}{c} B_{-}\\C_{-} \end{array} \right)\nonumber\\
\text{or,}\hspace{0.2in}\left(\begin{array}{cc} A_{-} & D_{-}\\ \end{array}\right)\left(\begin{array}{c} B_{-}\\C_{-} \end{array} \right)&=&\left(\begin{array}{cc} B_{-} & C_{-}\\ \end{array}\right)S^{\dagger}(V_{odd})S(-V_{odd})\left(\begin{array}{c} A_{-}\\D_{-} \end{array} \right)\nonumber\\
\text{and hence,}\hspace{0.2in} S^{\dagger}\left(V_{odd}\right)S\left(-V_{odd}\right)&=&\textit{I}, \label{Duality}
\end{eqnarray}
where the relation between the coefficients has been used. The final result is the 
\textit{Duality} condition already well-appreciated in the optical analogues of $\cal{PT}$-symmetric systems \cite{Dual}. 
Hermiticity, and subsequent unitarity, of the system is ensured for $V_{odd}\rightarrow 0$. This result is 
linked with the fact that $V_{odd}\rightarrow -V_{odd}$, essentially is the complex conjugation. In the above 
derivation, the asymptotic behavior of the $\cal{PT}$-symmetric system being Hermitian, has been utilized extensively. 
Also, the free particle solution having a unique momentum is a key fact here. It is to be noted that in deriving Eq.\ref{Duality},
nowhere the fact was utilized that the system is $\cal{PT}$-symmetric. The above is true for any complex potential. However, 
relations obtained in Ref.\cite{Dual} for elements of the $S$-matrix cannot be obtained here due to the aforementioned lack of
a suitable inner product, particularly when $\cal{PT}$-symmetry is preserved. 
\paragraph*{\bf The in-out interpretation:} Till now in this section, the usual asymptotic treatment for scattering has been carried out in terms of left
and right asymptotic states. However, in view of the non-locality of the scalar product, an  alternate but natural description can
be in terms of {\it initial}/{\it final} states, which are two incoming/outgoing plane waves from/to asymptotes on {\it both}
sides of the potential. This is further supported by the physical definition of $S$-matrix, yielding the final scattering state by
acting upon the initial one, as in Eq.\ref{def}. The identification that these are the allowed boundary conditions for scattering states, under 
$\cal{PT}$-symmetry, will be made in the next section. With this realization, we have 
$\psi_{in}(x)\equiv Ae^{ikx}+De^{-ikx}$ and $\psi_{out}(x)\equiv Ce^{ikx}+Be^{-ikx}$ respectively. Now, by equating
the fluxes (Eq.\ref{cont}), we have the $S$-matrix satisfying,

\begin{equation}
S^{\dagger}\left(
\begin{array}{cc} 0 & -1\\1 & 0
\end{array} \right)S=\left(
\begin{array}{cc} 0 & -1\\1 & 0
\end{array} \right),\label{corr}
\end{equation}
which is precisely the pseudo-Hermiticity condition $S^{\dagger}\eta=S^{-1}\eta$ \cite{Das}, with the norm operator
identified as $\eta=\left(\begin{array}{cc} 0 & -1\\1 & 0\end{array} \right)$. The unitarity of the $S$-matrix is restored
for $\eta=1$, expectedly, yielding back the Hermitian system and corresponding norm. It is to be noted that the pseudo-Hermiticity
obtained here is only for the asymptotic states, and is not established for all the states, and hence for the system itself.
However, incorporating the physical meaning of the non-local scalar for the choice of the scattering states leads to definite
conclusions, which will result in specific boundary conditions, obtained in the next section. 
\paragraph*{} Interestingly, in Hermitian quantum mechanics both left-right and in-out labeling of scattering states lead to  
same properties of the $S$-matrix and related boundary conditions \cite{Merzbacher}, though the second one is more physical.
This is because asymptotic states have definite energy corresponding to a unitary time evolution, which does {\it not} appear
in the local {\it stationary} scalar $\psi^*(x,t)\psi(x,t)$. Thus, whether or not a state is made out of simultaneous 
plane-wave components, does not make any difference. On the other hand, in $\cal{PT}$-symmetric systems, the scalar 
$\psi^*(-x,t)\psi(x,t)$ is both non-stationary (for scattering states) and {\it non-local}, imposing physical difference between
the two aforementioned labellings, picking out the in-out labeling for scattering states to be the observable one. This non-locality
is the central physical feature of such systems, leading to specific boundary conditions for scattering, which have been observed in
physical systems. It also leads to pseudo-Hermiticity, suggesting towards a proper norm. However,
the left-right choice can still be considered for mathematical purposes, especially for comparision with classical analogues of 
$\cal{PT}$-symmetric systems which are asymptotically Hermitian.

\section{Constrained Boundary Conditions: Bound and scattering states} The equation of continuity can also be utilized to study
$\cal{PT}$-symmetric systems, as the inherent symmetry of the system is incorporated within it. From Eq.\ref{cont}, if
$\cal{PT}$-symmetry is unbroken, \textit{i.e.}, if $\psi^{*}(-x,t)\equiv\psi(x,t)$, the `current' itself vanishes:

\begin{equation}
\frac{\partial}{\partial t}\psi^2(x,t)=0. \label{stationary}
\end{equation} 
Although the wave-function can still be complex in general, it is {\it explicitly} time-independent and hence, physically corresponds 
to a stationary state. This contradicts with the fact that $\cal{PT}$-symmetric states have {\it real} finite energy eigenvalues
\cite{Bender,Guo}. In Hermitian systems, a state is stationary modulo the unitary time evolution $\exp(-iEt)$. Similarly, $\cal{PT}$-symmetry
of an energy eigenfunction is to be defined modulo the same factor. When $E$ is real, then $\psi^*(-x,t)\psi(x,t)\equiv\psi^*(-x)\psi(x)$.
Further, as the `current' identically vanishes for $\cal{PT}$-symmetric states, they also are the {\it bound} states.
\paragraph*{} When $E$ is complex, the continuity equation becomes,

\begin{equation} 
\frac{\hbar^2}{2m}\frac{\partial}{\partial x}\left(\psi(x,t) \frac{\partial}{\partial x} {\psi}^{*}(-x,t)-{\psi}^{*}(-x,t)\frac{\partial}{\partial x} \psi(x,t)\right)=2E_{\text{im}}\left({\psi}^{*}(-x,t)\psi(x,t)\right),\label{im}
\end{equation}
yielding a non-vanishing current. This is the case of spontaneously broken $\cal{PT}$-symmetry, with complex-conjugate pairs of
eigenvalues \cite{Bender}. Upon $\cal{PT}$-transformation of the time-independent Schr\"odinger equation, it is seen that if the
energy eigenvalue is complex, then its complex-conjugate is also an eigenvalue. The corresponding eigenfunctions are related
through $\cal{PT}$-transformation, as there is no degeneracy in low dimention. Following the earlier arguments, these states are non-stationary,
and physically correspond to gain/decay \cite{Guo}.
\paragraph*{•} We now point out the constraints on boundary conditions, imposed by the conserved correlation. The critical
observation from Eq.\ref{flux} is that, as $A$ and $D$ are the respective amplitudes of the fluxes from $\mp\infty$, 
the absence of either, to begin with, makes the two amplitudes on the other side complex-conjugates. The
outgoing/incoming flux actually vanishes, if either of the concerned coefficients is zero. Moreover, as only the cross-terms appear in 
Eq.\ref{flux}, incident, reflected or transmitted fluxes are not intuitively separable. Further, the norm operator $\eta$ for such
systems is necessarily stationary \cite{Rivers}. Therefore, on physical
grounds, the non-local scalar for scattering states cannot be stationary, as the corresponding `current' must not vanish. Thus, from
Eq.\ref{cont}, the corresponding eigenfunction must be non-trivially time-dependent, in addition to the `energy exponent'. Also, it
\textit{cannot} be $\cal{PT}$-symmetric, or of any other form which makes the current vanish. This conclusion excludes scattering 
solutions like superpositions of pure $\cal{PT}$-symmetric/anti-symmetric functions, specifically plane waves with \textit{real} or 
\textit{pure imaginary} coefficients. It also cannot be a \textit{single} plane wave. Therefore, particular
boundary conditions, \textit{e.g.}, only incoming flux in any one side (left or right) of the potential, are automatically ruled out. One
can have a situation like incidence from left, resulting into reflection back, but \textit{no} transmission. Additionally, as the scattering
states are not $\cal{PT}$-symmetric, the corresponding eigenvalues cannot be real, and this physically means absorption/emission.
\paragraph*{•} The allowed scattering states correspond to incidence from and emission to both directions, which is precisely the
case for arriving at Eq.\ref{corr}, satisfying pseudo-Hermiticity in the process, with {\it complex} amplitudes. As the `current'
vanishes, wave-function only in one side cannot exist, thereby cannot be a scattering state. This condition was recently realized experimentally in 
the CPA, or anti-laser \cite{Cao}. Two coherent beams of laser were incident on a sample with an optical profile respecting $\cal{T}$-symmetry, 
which when unbroken, both reflection ($\Re$) and transmission ($\Im$) amplitudes were observed to vanish.
This was later shown to be a special case of the $\cal{PT}$ 
CPA laser \cite{Longhi}, which can generate stimulated emission, while shone with coherent radiation under suitable boundary 
conditions. It can also completely absorb that radiation for appropriate amplitude and phase relationship, which precisely is the CPA 
system. As coherent radiation is essentially classical in nature, the evaluation of $\Re$ and $\Im$ is straightforward. Here, we have a 
quantum mechanical analogue, utilizing plane waves instead of coherent radiation. This is analogous to the treatment of \cite{Longhi}, 
where plane waves are considered, representing individual Fourier components of laser. $\cal{PT}$-symmetry results in specific 
relations between the transfer matrix ($M$) elements, subsequently making the material a perfect absorber or emitter for suitable 
boundary conditions. For quantum systems, an extra input, the well-defined norm, is necessary to physically deal with matrix elements. 
Still, from our study, the equivalence is obvious between the boundary conditions. As is evident, the experimental realization of the 
$\cal{PT}$ CPA laser will shed much light on the structure of an appropriate inner product for $\cal{PT}$-symmetric systems. 

\section{Example: Scattering by $\cal{PT}$-symmetric Scarf-II potential}
As a demonstration of the above conclusions, we consider  $\cal{PT}$-symmetric {\it complexified} 1-D Scarf-II potential, 

\begin{equation}
V(x)= A^{2}-\left(A(A+\alpha)+B^{2}\right)\frac{1}{\cosh^{2}(\alpha x)}+iB(2A+\alpha)\frac{\tanh(\alpha x)}{\cosh(\alpha x)},\label{Example}
\end{equation}
obtained from the real counterpart, by suitable complexification \cite{AP}}. It asymptotically approaches
a real constant, allowing the scattering states to be plane-waves. This potential is exactly solvable, allowing an algebraic treatment 
under supersymmetric (SUSY) quantum mechanics \cite{susy}, with the superpotential, $W(x)=A\tanh(\alpha x)+iB/\cosh(\alpha x)$.
\paragraph*{} The asymptotic analysis leads to the transmission and reflection coefficients, respectively, as,

\begin{eqnarray}
\Im\left(k,\frac{A}{\alpha},\frac{iB}{\alpha}\right)&=&\frac{\Gamma \left[-A/{\alpha}-ik/{\alpha}\right]\Gamma \left[1+A/{\alpha}-ik/{\alpha}\right]\Gamma \left[\frac{1}{2}-B/{\alpha}-ik/{\alpha}\right]\Gamma \left[\frac{1}{2}+B/{\alpha}-ik/{\alpha}\right]}{\Gamma \left(-ik/{\alpha}\right)\Gamma \left[1+ik/{\alpha}\right]{\Gamma}^{2}\left[\frac{1}{2}-ik/{\alpha}\right]}~~~\text{and},\nonumber\\
\Re\left(k,\frac{A}{\alpha},\frac{iB}{\alpha}\right)&=&i\Im\left(k,\frac{A}{\alpha},\frac{iB}{\alpha}\right)\left[\frac{\cos(\pi A/{\alpha})\sin(\pi B/{\alpha})}{\cosh(\pi k/{\alpha})}+\frac{\sinh(\pi A/{\alpha})\cos(\pi B/{\alpha})}{\sinh(\pi k/{\alpha})}\right].\label{pole}
\end{eqnarray}
Here $k=\frac{\alpha}{i}\left(n-A/{\alpha}\right)$ is the asymptotic momentum and $n$ is the label of the corresponding normalized
eigenstate. Subsequently, under the Dirac-von-Neuman scalar product,

\begin{equation}
|\Re|^{2}+|\Im|^{2}=1+\left[\frac{2\cos^{2}(\pi A/{\alpha})\sin^{2}(\pi B/{\alpha})\sinh^{2}(\pi k/{\alpha})+\sin(2\pi A/{\alpha})\sin(2\pi B/{\alpha})\sinh(2\pi k/{\alpha})}{\left(\sinh^{2}(\pi k/{\alpha})+\sin^{2}(\pi A/{\alpha})\cos^{2}(\pi B/{\alpha})\right)\cosh^{2}(\pi k/{\alpha})-\cos^{2}(\pi A/{\alpha})\sin^{2}(\pi B/{\alpha})\sinh^{2}(\pi k/{\alpha})}\right]. \label{conservation}
\end{equation}
The flux is not conserved, owing to the imaginary part of the potential, causing absorption or emission. This fails 
to incorporate the unbroken $\cal{PT}$-symmetry phase, which has been experimentally 
established \cite{Guo}. The deviation term in the square bracket does not vanish for preserved $\cal{PT}$-symmetry 
\cite{AP}, which was also known earlier \cite{MP}. It does vanish for $B\rightarrow\pm iB$, yielding back the Hermitian system.
\paragraph*{•} In absence of a definite norm, the direct verification of flux conservation for a generic $\cal{PT}$-symmetric system is ambiguous,
and $|\Im|^{2}$ and $|\Re|^{2}$ need to be re-defined suitably. Owing to the realness of the discrete spectrum for unbroken $\cal{PT}$-symmetry, it is 
expected that the modified norm can be conserved. However, we have demonstrated that the characteristic scalar product of such systems is 
subjected to a natural non-local interpretation, thus altering the allowed boundary conditions altogether. We verify them explicitly in this example,
following the treatment for its real counterpart \cite{Khare1}. The corresponding Schr\"odinger equation has two independent solutions, which
asymptotically have the forms, 

\begin{equation}
F_{1,2}(x;A,B,{\alpha},k)\longrightarrow \left\{\begin{array}{rl}{A_{1,2}\exp(ikx)+B_{1,2}\exp(-ikx)} & \text{if } x\longrightarrow -\infty \\
{C_{1,2}\exp(ikx)+D_{1,2}\exp(-ikx)} & \text{if } x\longrightarrow \infty \end{array} \right. ,
\end{equation}
where,

\begin{eqnarray}
A_1&=&\frac{\Gamma\left(-B/{\alpha}-A/{\alpha}+\frac{1}{2}\right)\Gamma\left(-2ik/{\alpha}\right)}{\Gamma\left(-A/{\alpha}-ik/{\alpha}\right)\Gamma\left(-B/{\alpha}+\frac{1}{2}-ik/{\alpha}\right)}e^{\pi\left(k/{\alpha}+iB/{\alpha}+iA/{\alpha}\right)}-A/{\alpha}+2ik/{\alpha},\nonumber\\
B_1&=&\frac{\Gamma\left(-B/{\alpha}-A/{\alpha}+\frac{1}{2}\right)\Gamma\left(2ik/{\alpha}\right)}{\Gamma\left(-A/{\alpha}+ik/{\alpha}\right)\Gamma\left(-B/{\alpha}+\frac{1}{2}+ik/{\alpha}\right)}e^{\pi\left(-k/{\alpha}+iB/{\alpha}+iA/{\alpha}\right)}-A/{\alpha}-2ik/{\alpha},\nonumber\\
C_1&=&\frac{\Gamma\left(-B/{\alpha}-A/{\alpha}+\frac{1}{2}\right)\Gamma\left(2ik/{\alpha}\right)}{\Gamma\left(-A/{\alpha}+ik/{\alpha}\right)\Gamma\left(-B/{\alpha}+\frac{1}{2}+ik/{\alpha}\right)}e^{\frac{\pi}{2}\left(k/{\alpha}-iB/{\alpha}-iA/{\alpha}\right)}-A/{\alpha}-2ik/{\alpha},\nonumber\\
D_1&=&\frac{\Gamma\left(-B/{\alpha}-A/{\alpha}+\frac{1}{2}\right)\Gamma\left(-2ik/{\alpha}\right)}{\Gamma\left(-A/{\alpha}-ik/{\alpha}\right)\Gamma\left(-B/{\alpha}+\frac{1}{2}-ik/{\alpha}\right)}e^{\frac{\pi}{2}\left(-k/{\alpha}-iB/{\alpha}-iA/{\alpha}\right)}-A/{\alpha}+2ik/{\alpha},\nonumber\\
A_2&=&-i\frac{\Gamma\left(\frac{3}{2}+A/{\alpha}+B/{\alpha}\right)\Gamma\left(-2ik/{\alpha}\right)}{\Gamma\left(\frac{1}{2}+B/{\alpha}+ik/{\alpha}\right)\Gamma\left(1+A/{\alpha}-1k/{\alpha}\right)}e^{\pi\left(k/{\alpha}-iB/{\alpha}-iA/{\alpha}\right)}-A/{\alpha}+2ik/{\alpha},\nonumber\\
B_2&=&-i\frac{\Gamma\left(\frac{3}{2}+A/{\alpha}+B/{\alpha}\right)\Gamma\left(-2ik/{\alpha}\right)}{\Gamma\left(\frac{1}{2}+B/{\alpha}+ik/{\alpha}\right)\Gamma\left(1+A/{\alpha}+1k/{\alpha}\right)}e^{\pi\left(k/{\alpha}+iB/{\alpha}+
iA/{\alpha}\right)}-A/{\alpha}-2ik/{\alpha},\nonumber\\
C_2&=&i\frac{\Gamma\left(\frac{3}{2}+A/{\alpha}+B/{\alpha}\right)\Gamma\left(-2ik/{\alpha}\right)}{\Gamma\left(\frac{1}{2}+B/{\alpha}+ik/{\alpha}\right)\Gamma\left(1+A/{\alpha}+1k/{\alpha}\right)}e^{\frac{\pi}{2}\left(-k/{\alpha}+iB/{\alpha}+
iA/{\alpha}\right)}-A/{\alpha}-2ik/{\alpha},\nonumber\\
D_2&=&i\frac{\Gamma\left(\frac{3}{2}+A/{\alpha}+B/{\alpha}\right)\Gamma\left(-2ik/{\alpha}\right)}{\Gamma\left(\frac{1}{2}+B/{\alpha}+ik/{\alpha}\right)\Gamma\left(1+A/{\alpha}-1k/{\alpha}\right)}e^{\frac{\pi}{2}\left(-k/{\alpha}+iB/{\alpha}+iA/{\alpha}\right)}-A/{\alpha}+2ik/{\alpha}.
\end{eqnarray}
These coefficients are all complex for arbitrary momentum $k$, and do not vanish in general. The phase factor in each 
of them carries a term linear in $iB$, the parameter signifying $\cal{PT}$-symmetry, ensuring the overall complexity 
of the coefficients. This is in accordance with the boundary conditions obtained in the previous section for scattering states,
that plane waves from both $x=\pm\infty$, with complex coefficients, must constitute asymptotic states. 
\paragraph*{•} Clearly, the system is asymptotically Hermitian, and the flux attenuation/enhancement takes place locally. 
Thus the conclusion from Eq.\ref{conservation}, which is asymptotically valid, is justified. Despite the system being
asymptotically Hermitian, the scattered particle `carries' the memory of the local symmetry of the Hamiltonian in terms of
the constraints on the coefficients, which restricts our choice. Thus we can justify Eq.\ref{flux}, and interpret the asymptotic
behavior of a $\cal{PT}$-symmetric system as the manifestation of non-stationarity of a scattering state, subjected to the
intrinsic symmetry of the system. 

\section{Conclusion} 
In conclusion, suitable continuity equation can be constructed and utilized for $\cal{PT}$-symmetric systems to obtain informations 
about the nature of scattering and bound states. It results in a conserved non-local scalar product, necessitating the presence of both incoming and 
outgoing states for the asymptotic case of scattering. Further, instead of a local probability density, non-local correlation dictates the 
structure of bound, resonance and scattering states. Corresponding boundary conditions have exact analogues for $\cal{PT}$ CPA laser and 
other optical systems. The lack of a local norm for the generic scattering restricts the proper extraction of reflection and transmission 
coefficients, further illuminating the inherent non-locality of such systems. However, classical analogues of such systems are understood,
especially optical ones \cite{Longhi}, which bypass these difficulties, and can have various practical use as switches and detectors. Further,
non-linear quantum mechanical $\cal{PT}$-symmetric systems \cite{NLD} can yield novel conditions for stability of solutions in light of the unique boundary
conditions proposed here.

\end{document}